\documentclass[conference]{IEEEtran}
\IEEEoverridecommandlockouts

\usepackage{cite}
\usepackage{amsmath,amssymb,amsfonts}
\usepackage{graphicx}
\usepackage{textcomp}
\usepackage{xcolor}
\usepackage{booktabs}
\usepackage{url}
\usepackage{listings}

\def\BibTeX{{\rm B\kern-.05em{\sc i\kern-.025em b}\kern-.08em
		T\kern-.1667em\lower.7ex\hbox{E}\kern-.125emX}}

\begin{document}
	
	\title{Can LLMs Design Video Coding Tools? \\ A Case Study on Planar Mode}
	
	\author{
		\IEEEauthorblockN{
			Yingwen Zhang$^{1}$,
			Meng Wang$^{2}$,
			Liqiang He$^{1}$,
			Shiqi Wang$^{1\ddagger}$
			%    Shiqi Wang$^{1}$
		}
		\IEEEauthorblockA{
			$^{1}$Department of Computer Science, City University of Hong Kong\\
			$^{2}$School of Data Science, Lingnan University
		}
		\thanks{$\ddagger$ Corresponding author}
	}
	
	\IEEEaftertitletext{\vspace{-1.5em}}
	
	\maketitle
	
	\begin{abstract}
		This paper explores whether large language models (LLMs) can design video coding tools, a highly challenging task due to the intricate algorithmic coupling of tool modifications. In particular, we present an empirical case study on the Planar mode, a long-standing intra prediction tool in video coding standards. Our experiments operate within a generation-and-evaluation loop, with the LLM generating new Planar predictors, encoder trials evaluating their coding performance, and the LLM re-generating refined implementations based on the evaluation feedback. We first examine directly replacing the default Planar mode in the Fraunhofer Versatile Video Encoder (VVenC) under its \textit{faster} preset. Experimental results demonstrate that the LLM-generated mode can outperform the conventional Planar mode on this lightweight toolset, achieving 0.18\% bitrate savings with 0.4\% complexity overhead on the standard benchmark. We further extend our evaluation to the Enhanced Compression Model (ECM). Leveraging newly introduced directional Planar modes, we investigate two integration strategies: directly replacing them, and introducing the LLM-generated predictor as an additional prediction mode with new syntax elements. The empirical results suggest that both strategies can yield coding gains under a constrained low-resolution setting. Overall, this study offers preliminary evidence and practical insights, highlighting both the potential and open challenges of LLM-based coding tool design. 
		%VVenC Planar test codes are available at: 
	\end{abstract}
	
	
	\section{Introduction}
	Large language models (LLMs) have recently expanded from general question-answering tasks to more structured applications in automatic algorithm design~\cite{romeraparedes2024funsearch,liu2024eoh,novikov2025alphaevolve}. In video coding, this paradigm has shown promising results in quantization parameter (QP) allocation~\cite{he2026llmdrivenqp}. Typically, an evolutionary generation-and-evaluation loop is constructed, where the LLM acts as a designer to iteratively propose QP decision rules, actual encoder executions evaluate these rules to compute BD-rate~\cite{bjontegaard2001calculation} performance, and the resulting feedback guides the subsequent refinement of the algorithm. This line of research demonstrates that LLMs, when guided by proper encoding feedback, can produce effective code-level solutions for well-defined encoder optimization tasks. A critical next question is whether this approach can be extended to designing video coding tools themselves.
	
	%This extension is essential, as coding tools are the primary drivers of video coding standard evolution. 
	From Advanced Video Coding (AVC)~\cite{wiegand2003overview} to Versatile Video Coding (VVC)~\cite{bross2021overview} and toward the emerging next-generation standard~\cite{JVET-AI0247}, coding gains have largely resulted from the iterative proposal, validation, and refinement of coding tools. Meanwhile, the coding tool development remains heavily expert-driven: domain experts conceive an idea, integrate it into reference software, evaluate its coding efficiency and complexity, and iteratively improve it before standardization discussion. While effective, the explored design space is inevitably limited by human domain knowledge. LLMs provide a new opportunity to expand this search space. However, unlike a QP allocation rule~\cite{he2026llmdrivenqp} applied at the pre-encoding stage, a coding tool is deeply embedded in the core codec pipeline. Consequently, its modifications introduce an intricate algorithmic coupling with surrounding modules. For example, modifications to a prediction tool may propagate a cascading impact to downstream stages such as transform, quantization, and rate-distortion optimization (RDO). Furthermore, any logic adjustment must carefully account for syntax signaling overhead and strict encoder-decoder matching. These stringent requirements make coding tool design significantly more challenging for LLMs.
	
	In this paper, we conduct a case study on the Planar mode~\cite{lainema2010intra}, a foundational intra prediction tool across multiple video coding standards. Its basic form is a linear interpolation from reconstructed boundary samples, making it a simple and interpretable starting point for investigating LLM-based tool design. We ask a narrow question: under existing codec syntax and interface constraints, can an LLM generate a Planar function that outperforms the hand-crafted baseline in certain configurations? To answer this, we explore the direct mode replacement in the Fraunhofer Versatile Video Encoder (VVenC)~\cite{VVenC} \textit{faster} preset, and then further evaluate both direct replacement and new-mode insertion in the Enhanced Compression Model (ECM)~\cite{jvet2025ecm19} based on the directional Planar mode~\cite{jvet2023planarhv} implementation. This work represents a first step toward examining where LLM-based coding tool design shows promise and what remains open.
	
	\section{Methodology}
	\subsection{Preliminaries: Planar Prediction}
	Planar prediction~\cite{lainema2010intra} was introduced to approximate smoothly varying two-dimensional surfaces. Rather than explicitly signaling spatial control points, it interpolates the lower-right texture region by fully leveraging the reconstructed boundary samples. Formally, given a coding block of width $W$ and height $H$, the prediction value $\hat{P}(x,y)$ within $0 \le x < W$ and $0 \le y < H$ is defined as a linear combination of horizontal and vertical interpolations. While the left boundary $L(-1,y)$ and top boundary $T(x,-1)$ represent readily available reconstructed samples, the virtual bottom-right references are extrapolated from the bottom-left corner $L(-1,H)$ and top-right corner $T(W,-1)$, respectively. The resulting mathematical formulation is expressed as
	\begin{equation}
		\hat{P}(x,y) = \frac{HF_{\mathrm{h}}(x,y) + WF_{\mathrm{v}}(x,y)}{2WH},
		\label{eq:planar1}
	\end{equation}
	where the horizontal and vertical interpolation components, $F_{\mathrm{h}}(x,y)$ and $F_{\mathrm{v}}(x,y)$, are defined as:
	\begin{equation}
		\begin{aligned}
			F_{\mathrm{h}}(x,y) &= (W-1-x)L(-1,y) + (x+1)T(W,-1), \\
			F_{\mathrm{v}}(x,y) &= (H-1-y)T(x,-1) + (y+1)L(-1,H).
		\end{aligned}
		\label{eq:planar_components}
	\end{equation}
	Owing to its high coding efficiency for homogeneous regions, Planar prediction serves as a foundational intra tool and is commonly prioritized in the construction of the most-probable-mode (MPM) list~\cite{jvet2019unifiedmpm}.  
	
	Recent ECM development further extends this family with directional Planar variants that retain only one interpolation direction~\cite{jvet2023planarhv}. Let $\hat{P}_{\mathrm{h}}(x,y)$ and $\hat{P}_{\mathrm{v}}(x,y)$ denote the predicted sample values of the horizontal and vertical directional Planar modes, respectively. They can be computed as
	\begin{equation}
		\begin{aligned}
			\hat{P}_{\mathrm{h}}(x,y) &= \frac{1}{W} \cdot F_{\mathrm{h}}(x,y), \\
			\hat{P}_{\mathrm{v}}(x,y) &= \frac{1}{H} \cdot F_{\mathrm{v}}(x,y).
		\end{aligned}
		\label{eq:directional_planar}
	\end{equation}
	These two variants are pre-selected by the Hadamard cost~\cite{Hadamard} and then inserted into the full RDO list when competitive. If a directional Planar mode is used, it is signaled through the Planar branch of the MPM syntax, with an index distinguishing the original, horizontal, and vertical variants. The transform side is also adapted to the directional bias: the horizontal and vertical Planar variants use the corresponding horizontal and vertical intra modes to derive the transform kernels~\cite{jvet2023planarhv}.
	
	\subsection{LLM-Based Planar Mode Design}
	The core of the Planar-family mode is a prediction function $f(\cdot)$ that maps reconstructed boundaries to a two-dimensional prediction block. In this study, we examine whether such prediction functions can be automatically designed by LLMs. Our workflow follows an iterative generation-and-evaluation loop: given a collection of historical candidates and their coding feedback, the LLM generates novel, codec-compliant C++ code of $f(\cdot)$ alongside the underlying design ideas. Importantly, the search target of LLMs is deliberately limited to the core predictor body, while the surrounding codebase such as reference preparation remains unchanged. Each generated candidate is then syntactically validated, compiled into the codec, and evaluated through actual encoding runs. The measured coding performance is subsequently fed back into the next prompting iteration, allowing the workflow to accumulate this empirical history and iteratively search for better predictor formulations.
	
	Herein, to ensure that the generated Planar predictors remain structurally lightweight and compatible with codec implementation norms, we simply enforce these operational boundaries through prompt engineering. Each prompt is structured around five core components. First, the task description explicitly defines the target objective as designing novel Planar prediction mechanisms. Second, the historical feedback component introduces parent candidates from prior iterations; crucially, each parent candidate encapsulates both its conceptual design idea and its corresponding C++ code, paired with its sequence-level BD-rate performance. Third, the strategy-specific instruction implements two evolutionary search strategies~\cite{liu2024eoh}: exploration instructions direct the LLM to synthesize predictors that are structurally distinct from existing parent candidates, whereas modification instructions guide the LLM to perform localized structural or numeric variations on a single parent candidate. Fourth, the interface specification strictly defines the accessible input variables and mandates that the generated code outputs the prediction block. Fifth, the validity constraints enforce that the proposed predictors are decoder-reproducible and computationally practical. Candidates that fail syntactic parsing, interface compliance, compilation, or decoding are rejected and not retained for the next iteration. Fig.~\ref{fig:vvenc-planar-prompt-example} shows a compact prompt example for the VVenC Planar design used in our direct-replacement experiments.
	
	\begin{figure}[t]
		\centering
		\setlength{\fboxsep}{1.1mm}
		\fbox{%
			\begin{minipage}{0.95\columnwidth}
				\scriptsize
				\textbf{\textcolor{blue}{Task description.}}
				Optimize the prediction rule used by the original Planar mode in VVenC. Given the prepared top and left reference lines for the current block, generate a deterministic intra prediction rule that fills the whole block. Find practical and stable coding-gain improvements under strict encoder-decoder matching.
				
				\smallskip
				\textbf{\textcolor{green!50!black}{Historical feedback.}}
				(Inserted multiple parent predictors, including a short design idea, their raw C++ body, and per-sequence BD-rate feedback).
				
				\smallskip
				\textbf{\textcolor{orange}{Strategy-specific instruction.}}
				Create a new predictor that is structurally different from the given ones. Keep the new design in the same or lower complexity class as the parents. First, write exactly one sentence inside \texttt{$<$idea$>$...$<$/idea$>$} that describes the new algorithm at a high level. Next, write exactly the code block inside \texttt{\textasciigrave\textasciigrave\textasciigrave cpp\textasciigrave\textasciigrave\textasciigrave}.
				
				\smallskip
				\textbf{\textcolor{blue!70!black}{Interface specification.}}
				The generated C++ body is inserted inside \texttt{xPredIntraPlanar\_Core}. The surrounding function has already prepared \texttt{width}, \texttt{height}, \texttt{stride}, \texttt{pred}, \texttt{top[]}, and \texttt{left[]}. The generated code must write every sample through \texttt{pred[y * stride + x]}.
				
				\smallskip
				\textbf{\textcolor{magenta}{Validity constraints.}}
				Use deterministic integer arithmetic and valid boundary indexing. Keep the main prediction loop lightweight; statistics precomputation is allowed when useful. Do not use encoder-only state, randomness, dynamic allocation, I/O, preprocessor directives, or external helper logic.
		\end{minipage}}
		\caption{A compact (exploration) prompt example for VVenC Planar design.}
		\label{fig:vvenc-planar-prompt-example}
	\end{figure}
	
	After encoding, the generated predictors are merged into a historical candidate pool. We sort this pool according to an objective $\mathcal{J}$ and retain the top-$K$ candidates as parents for subsequent prompts. Since the average BD-rate $\bar{\mathcal{B}}(f)$ alone can favor predictors that improve a few sequences while degrading others, we add a worst-case-aware regularization term $\mathcal{R}(f)$:
	\begin{gather}
		\mathcal{J}(f)
		= \bar{\mathcal{B}}(f) + \lambda \mathcal{R}(f),
		\label{eq:llm_objective}
	\end{gather}
	where $\mathcal{R}(f)$ is the average BD-rate of the worst-$M$ sequences, and $\lambda$ controls the regularization strength. This objective is used for sorting and parent sampling. For each exploration prompt, we sample $N$ parent candidates based on ranking, where candidates with lower rank indices are assigned higher probability. In our experiment, we fix $K=16$, $M=2$, $N=8$, and generate 64 candidates for each iteration by repeating the exploration and modification prompts. The core LLM is DeepSeek-V4-Flash~\cite{deepseekai2026deepseekv4} with high thinking mode.
	
	%.\footnote{-ip 1 --GOPSize 1 --MCTF 0 --BIM 0 --IntraQPOffset 0}
	\section{Experiments}
	
	\begin{table}[t]
		%	\caption{Two evolved Planar predictors over VVenC-1.14.0 \textit{faster}.}
		\caption{Evolved Planar results (Anchor: Default VVenC-1.14.0 \textit{faster}).}
		\label{tab:vvenc-result}
		\centering
		%\footnotesize
		\setlength{\tabcolsep}{4.2pt}
		\begin{tabular}{lrrrr}
			\toprule
			& \multicolumn{2}{c}{$\lambda=2$} & \multicolumn{2}{c}{$\lambda=5$} \\
			\cmidrule(lr){2-3}\cmidrule(lr){4-5}
			Sequence & BD-rate & $\Delta$EncT & BD-rate & $\Delta$EncT \\
			\midrule
			Tango2 & $-0.14\%$ & $+0.9\%$ & $-0.05\%$ & $+5.9\%$ \\
			FoodMarket4 & $-0.15\%$ & $+0.6\%$ & $-0.03\%$ & $+5.9\%$ \\
			Campfire & $-0.23\%$ & $+0.8\%$ & $-0.16\%$ & $+4.4\%$ \\
			CatRobot & $-0.14\%$ & $+1.1\%$ & $-0.12\%$ & $+4.5\%$ \\
			DaylightRoad2 & $-0.07\%$ & $+0.7\%$ & $-0.10\%$ & $+5.0\%$ \\
			ParkRunning3 & $-0.12\%$ & $0.0\%$ & $-0.07\%$ & $+3.6\%$ \\
			\textbf{Class A Average} & $\mathbf{-0.14\%}$ & $\mathbf{+0.7\%}$ & $\mathbf{-0.09\%}$ & $\mathbf{+4.9\%}$ \\
			MarketPlace & $-0.16\%$ & $+0.9\%$ & $-0.13\%$ & $+5.2\%$ \\
			RitualDance & $-0.17\%$ & $+1.5\%$ & $-0.12\%$ & $+4.5\%$ \\
			Cactus & $-0.22\%$ & $+0.1\%$ & $-0.16\%$ & $+3.5\%$ \\
			BasketballDrive & $-0.16\%$ & $+1.2\%$ & $-0.13\%$ & $+3.8\%$ \\
			BQTerrace & $-0.17\%$ & $+0.3\%$ & $-0.20\%$ & $+3.8\%$ \\
			\textbf{Class B Average} & $\mathbf{-0.18\%}$ & $\mathbf{+0.8\%}$ & $\mathbf{-0.15\%}$ & $\mathbf{+4.2\%}$ \\
			BasketballDrill & $-0.22\%$ & $-0.4\%$ & $-0.21\%$ & $+4.0\%$ \\
			BQMall & $-0.11\%$ & $+0.4\%$ & $-0.16\%$ & $+4.6\%$ \\
			PartyScene & $-0.34\%$ & $-0.8\%$ & $-0.27\%$ & $+3.0\%$ \\
			RaceHorsesC & $-0.27\%$ & $0.0\%$ & $-0.22\%$ & $+4.0\%$ \\
			\textbf{Class C Average} & $\mathbf{-0.23\%}$ & $\mathbf{-0.2\%}$ & $\mathbf{-0.22\%}$ & $\mathbf{+3.9\%}$ \\
			BasketballPass & $-0.14\%$ & $+0.4\%$ & $-0.09\%$ & $+2.3\%$ \\
			BQSquare & $-0.27\%$ & $-0.2\%$ & $-0.27\%$ & $+2.3\%$ \\
			BlowingBubbles & $-0.32\%$ & $-0.1\%$ & $-0.27\%$ & $+2.5\%$ \\
			RaceHorses & $-0.23\%$ & $-0.2\%$ & $-0.20\%$ & $+2.4\%$ \\
			\textbf{Class D Average} & $\mathbf{-0.24\%}$ & $\mathbf{0.0\%}$ & $\mathbf{-0.21\%}$ & $\mathbf{+2.4\%}$ \\
			FourPeople & $-0.06\%$ & $+1.4\%$ & $-0.18\%$ & $+4.7\%$ \\
			Johnny & $-0.22\%$ & $+0.9\%$ & $-0.19\%$ & $+4.4\%$ \\
			KristenAndSara & $-0.04\%$ & $+0.4\%$ & $-0.18\%$ & $+4.9\%$ \\
			\textbf{Class E Average} & $\mathbf{-0.11\%}$ & $\mathbf{+0.9\%}$ & $\mathbf{-0.18\%}$ & $\mathbf{+4.7\%}$ \\
			ArenaOfValor & $-0.20\%$ & $+0.4\%$ & $-0.24\%$ & $+4.3\%$ \\
			BasketballDrillText & $-0.22\%$ & $+0.7\%$ & $-0.23\%$ & $+3.7\%$ \\
			SlideEditing & $-0.08\%$ & $+0.7\%$ & $-0.04\%$ & $+1.5\%$ \\
			SlideShow & $-0.16\%$ & $-0.3\%$ & $-0.13\%$ & $+2.0\%$ \\
			\textbf{Class F Average} & $\mathbf{-0.17\%}$ & $\mathbf{+0.4\%}$ & $\mathbf{-0.16\%}$ & $\mathbf{+2.9\%}$ \\
			\midrule
			\textbf{All Average} & $\mathbf{-0.18\%}$ & $\mathbf{+0.4\%}$ & $\mathbf{-0.16\%}$ & $\mathbf{+3.9\%}$ \\
			\bottomrule
		\end{tabular}
	\end{table}
	
	\subsection{VVenC Mode Replacement}
	We first evaluate the direct Planar replacement in VVenC-1.14.0~\cite{VVenC} under the lightweight \textit{faster} preset, where most of the VVC intra tools are disabled. Our experiment directly reuses the existing Planar syntax, RDO process, and prepared reference samples in VVenC. Only the core Planar predictor operating on these references is replaced by an LLM-generated one. 
	As the all-intra (AI) configuration is not provided in VVenC, we modify the random access configuration to match the common test conditions (CTC). We turn on the Wavefront Parallel Processing (WPP) with four threads. The YUV-PSNR ratio is 6:1:1. In training, we use the default encoder as the baseline and evaluate the performance of the candidates on the first frame of the 22 natural sequences from CTC (Classes A to E). After 20 iterations of training, we test the candidates of the last iteration on full-length CTC sequences. 
	
	In Table~\ref{tab:vvenc-result}, coding performance of two evolved predictors at $\lambda=2$ and $\lambda=5$ is demonstrated. Both predictors achieve consistent bitrate savings over the VVenC anchor, with average BD-rate reductions of $0.18\%$ and $0.16\%$, respectively. The $\lambda=2$ candidate yields slightly larger coding gains with only a $0.4\%$ time increase, while the $\lambda=5$ candidate achieves a smaller BD-rate reduction. 
	%We also test a relaxed prompt setting in VVenC \textit{faster}, where the LLM is allowed to explore more complex functions. In this setting, the coding performance improves more noticeably, but the resulting functions exceed a practical complexity budget. This experiment separates two questions that are often coupled: whether the LLM can find a stronger predictor, and whether the predictor is deployable. The answer is mixed. Relaxing complexity suggests that the function space contains stronger candidates, and that the LLM can discover them. However, a coding tool must be simple enough to run repeatedly inside the encoder and decoder. Therefore, the relaxed setting is useful as an upper-bound exploration, but it does not by itself provide a practical coding tool.
	Interestingly, the two selected predictors correspond to two clearly different evolutionary directions. The $\lambda=2$ predictor improves Planar mainly through adaptive boundary filtering: It first computes the gradient of the top and left reference lines as
	\begin{equation}
		\begin{aligned}
			V_T &= \sum_{i=1}^{W}|T(i,-1)-T(i-1,-1)|,\\
			V_L &= \sum_{j=1}^{H}|L(-1,j)-L(-1,j-1)|,
		\end{aligned}
	\end{equation}  
	and then applies a three-tap smoothing filter with coefficients $[1,2,1]$ to both reference lines. If one side is much less smooth than the other, namely $V_T>5V_L/4$ or $V_L>5V_T/4$, the same filter is applied twice to that side. This gives
	\begin{equation}
		\tilde{T}=
		\begin{cases}
			\mathcal{S}(\mathcal{S}(T)), & V_T>5V_L/4,\\
			\mathcal{S}(T), & \mathrm{otherwise},
		\end{cases}
	\end{equation}
	and $\tilde{L}$ is obtained symmetrically. Subsequently, the predictor applies Eqn.~(\ref{eq:planar_components}) to $\tilde{T}$ and $\tilde{L}$ to obtain the two interpolation components $\tilde{F}_{\mathrm{h}}$ and $\tilde{F}_{\mathrm{v}}$. Finally, this predictor tunes the last averaging step by a gradient-dependent offset $\rho$:
	\begin{equation}
		\hat{P}_{\lambda=2}(x,y)
		=
		\frac{
			H\tilde{F}_{\mathrm{h}}(x,y)
			+
			W\tilde{F}_{\mathrm{v}}(x,y)
			+\rho}{2WH},
		\label{eq:evolved_planar_lambda2}
	\end{equation}
	\begin{equation}
		\text{where}\quad\rho=
		\mathrm{clip}\left(
		\frac{WH}{2}
		+
		\frac{16(V_T-V_L)}{V_T+V_L+32},
		0,\frac{WH}{2}
		\right),
	\end{equation}
	and $\mathrm{clip}(x,l,u)=\min(\max(x,l),u)$. Thus, this candidate keeps the planar philosophy, but suppresses local reference fluctuations before prediction and slightly shifts the final averaged surface according to the top-left variation imbalance.
	
	The $\lambda=5$ predictor instead changes how the Planar surface is formed. It first constructs a Planar-like surface using the same two components in Eqn.~(\ref{eq:planar_components}), but normalizes their sum by $W+H$ rather than using the original weights in Eqn.~(\ref{eq:planar1}):
	\begin{equation}
		P_{\mathrm{s}}(x,y)=
		\frac{F_{\mathrm{h}}(x,y)+F_{\mathrm{v}}(x,y)}{W+H}.
	\end{equation}
	It then computes the center-oriented references
	\begin{equation}
		\begin{aligned}
			C_T &= \frac{T(W/2-1,-1)+T(W/2,-1)}{2},\\
			C_L &= \frac{L(-1,H/2-1)+L(-1,H/2)}{2},
		\end{aligned}
	\end{equation}
	and blends the average $C=(C_T+C_L)/2$ with $P_{\mathrm{s}}(x,y)$ as
	\begin{equation}
		Q(x,y)=
		\frac{(32-a)P_{\mathrm{s}}(x,y)+aC}{32},
		\label{eq:evolved_planar_lambda5}
	\end{equation}
	where $a=8$ for $WH\le64$, $a=4$ for $WH\le256$, and $a=1$ otherwise. Thus, smaller blocks receive a stronger center-reference contribution. Finally, the predictor computes the global boundary mean $\bar{R}$ and boundary range $[R_{\min},R_{\max}]$. If $Q(x,y)$ falls inside this range, it is kept unchanged; otherwise, it is softly pulled toward $B=(C+\bar{R})/2$:
	\begin{equation}
		\hat{P}_{\lambda=5}(x,y)=
		\begin{cases}
			Q(x,y), & \mathrm{inside},\\
			(3Q(x,y)+B)/4, & \mathrm{otherwise}.
		\end{cases}
	\end{equation}
	%Compared with the $\lambda=2$ predictor, this rule does not filter the references. Instead, it regularizes the predicted surface itself by combining a reweighted Planar surface, a block-size-dependent center prior, and a boundary-range correction, which partially explains its different coding-complexity tradeoff in Table~\ref{tab:vvenc-result}.
	
	\begin{table}[t]
		%	\caption{Two evolved Planar predictors over ECM-19.1.}
		\caption{Evolved Planar results (Anchor: Default ECM-19.1).}
		\label{tab:preliminary-result}
		\centering
		%\footnotesize
		\setlength{\tabcolsep}{4.2pt}
		\begin{tabular}{lrrrr}
			\toprule
			416x240 & \multicolumn{2}{c}{Mode replacement} & \multicolumn{2}{c}{Mode insertion} \\
			resolution & BD-rate & $\Delta$EncT & BD-rate & $\Delta$EncT \\
			\midrule
			%		Tango2 & $-0.14\%$ & $+0.9\%$ & $-0.05\%$ & $+5.9\%$ \\
			%		FoodMarket4 & $-0.15\%$ & $+0.6\%$ & $-0.03\%$ & $+5.9\%$ \\
			%		Campfire & $-0.23\%$ & $+0.8\%$ & $-0.16\%$ & $+4.4\%$ \\
			%		CatRobot & $-0.14\%$ & $+1.1\%$ & $-0.12\%$ & $+4.5\%$ \\
			%		DaylightRoad2 & $-0.07\%$ & $+0.7\%$ & $-0.10\%$ & $+5.0\%$ \\
			%		ParkRunning3 & $-0.12\%$ & $0.0\%$ & $-0.07\%$ & $+3.6\%$ \\
			%		\textbf{Class A Average} & $\mathbf{-0.14\%}$ & $\mathbf{+0.7\%}$ & $\mathbf{-0.09\%}$ & $\mathbf{+4.9\%}$ \\
			%		MarketPlace & $-0.16\%$ & $+0.9\%$ & $-0.13\%$ & $+5.2\%$ \\
			%		RitualDance & $-0.17\%$ & $+1.5\%$ & $-0.12\%$ & $+4.5\%$ \\
			%		Cactus & $-0.22\%$ & $+0.1\%$ & $-0.16\%$ & $+3.5\%$ \\
			%		BasketballDrive & $-0.16\%$ & $+1.2\%$ & $-0.13\%$ & $+3.8\%$ \\
			%		BQTerrace & $-0.17\%$ & $+0.3\%$ & $-0.20\%$ & $+3.8\%$ \\
			%		\textbf{Class B Average} & $\mathbf{-0.18\%}$ & $\mathbf{+0.8\%}$ & $\mathbf{-0.15\%}$ & $\mathbf{+4.2\%}$ \\
			BasketballDrill & $-0.0020\%$ & $-1.33\%$ & $-0.0298\%$ & $+0.21\%$ \\
			BQMall & $-0.0488\%$ & $-0.52\%$ & $-0.0005\%$ & $-3.26\%$ \\
			PartyScene & $-0.0128\%$ & $+2.35\%$ & $-0.0184\%$ & $+2.48\%$ \\
			RaceHorsesC & $-0.0119\%$ & $+1.87\%$ & $-0.0162\%$ & $+3.45\%$ \\
			%		\textbf{Class C Average} & $\mathbf{-0.23\%}$ & $\mathbf{-0.2\%}$ & $\mathbf{-0.22\%}$ & $\mathbf{+3.9\%}$ \\
			BasketballPass & $-0.0141\%$ & $+1.55\%$ & $-0.0595\%$ & $+3.90\%$ \\
			BQSquare & $-0.0724\%$ & $-1.83\%$ & $-0.0163\%$ & $+0.03\%$ \\
			BlowingBubbles & $-0.0439\%$ & $-1.51\%$ & $-0.0401\%$ & $+0.43\%$ \\
			RaceHorses & $-0.0111\%$ & $+2.60\%$ & $-0.0723\%$ & $+0.33\%$ \\
			%		\textbf{Class D Average} & $\mathbf{-0.24\%}$ & $\mathbf{0.0\%}$ & $\mathbf{-0.21\%}$ & $\mathbf{+2.4\%}$ \\
			%		FourPeople & $-0.06\%$ & $+1.4\%$ & $-0.18\%$ & $+4.7\%$ \\
			%		Johnny & $-0.22\%$ & $+0.9\%$ & $-0.19\%$ & $+4.4\%$ \\
			%		KristenAndSara & $-0.04\%$ & $+0.4\%$ & $-0.18\%$ & $+4.9\%$ \\
			%		\textbf{Class E Average} & $\mathbf{-0.11\%}$ & $\mathbf{+0.9\%}$ & $\mathbf{-0.18\%}$ & $\mathbf{+4.7\%}$ \\
			%		ArenaOfValor & $-0.20\%$ & $+0.4\%$ & $-0.24\%$ & $+4.3\%$ \\
			%		BasketballDrillText & $-0.22\%$ & $+0.7\%$ & $-0.23\%$ & $+3.7\%$ \\
			%		SlideEditing & $-0.08\%$ & $+0.7\%$ & $-0.04\%$ & $+1.5\%$ \\
			%		SlideShow & $-0.16\%$ & $-0.3\%$ & $-0.13\%$ & $+2.0\%$ \\
			%		\textbf{Class F Average} & $\mathbf{-0.17\%}$ & $\mathbf{+0.4\%}$ & $\mathbf{-0.16\%}$ & $\mathbf{+2.9\%}$ \\
			%		\midrule
			\textbf{All Average} & $\mathbf{-0.0271\%}$ & $\mathbf{+0.42\%}$ & $\mathbf{-0.0316\%}$ & $\mathbf{+0.90\%}$ \\
			\bottomrule
		\end{tabular}
	\end{table}
	
	\subsection{ECM Mode Replacement and Insertion}
	We then extend the experiment to ECM-19.1~\cite{jvet2025ecm19}, the state-of-the-art reference platform with more competitive coding tools. Based on the two directional Planar modes~\cite{jvet2023planarhv}, we study two integration strategies: directly replacing the existing directional predictors and inserting a new Planar-like predictor as an additional mode. The replacement experiment follows the VVenC setting. For the insertion experiment, the generated mode is placed after the original Planar mode and before the two directional Planar modes, with its own entropy context. In the rough RDO stage, the directional-Planar candidate list is expanded to include the generated mode. For transform selection, the generated mode is treated as the original Planar.
	
	To keep the evaluation cost manageable, we restrict the experiment to eight 416$\times$240 sequences, including four Class D sequences and four Class C sequences downsampled to the same resolution. Training uses one frame per sequence, while testing reports the performance over ten frames. All ECM experiments run 30 iterations. The YUV-PSNR ratio is 10:1:1~\cite{jvet2026cccmclipping1011}. The testing results are shown in Table~\ref{tab:preliminary-result}. Under this specific setting, both integration strategies achieve consistent coding gains. Compared with the ECM anchor, directly replacing the two directional Planar predictors gives 0.0271\% BD-rate reduction with 0.42\% encoding-time increase, while inserting a new Planar-like mode gives 0.0316\% BD-rate reduction with 0.90\% encoding-time increase. Importantly, considering that the reported BD-rate of the directional Planar for Class D is around -0.05\%~\cite{jvet2023planarhv}, these improvements suggest that LLMs within a generation-and-evaluation loop can find meaningful Planar-family variants in a more saturated ECM setting. Since the experiment only covers eight low-resolution sequences, we treat these results as empirical observations rather than a concrete validation.
	
	Surprisingly, the evolved replacement predictor performs operations similar to those of the VVenC ($\lambda=2$): the two directional predictors still keep the one-direction interpolation structure, but each branch applies filtering to its one-sided boundary before prediction. The smoothed boundary is then conditionally blended with the original boundary sample according to the local gradient: relatively flat regions use more of the three-tap $[1,2,1]$ filtered reference, whereas sharp local transitions are mostly preserved. Thus, the replacement mode mainly modifies the stability of the boundary signal. The inserted mode is technically different. Although the implementation is still compact, its behavior is intricate with more conditioning and clipping operations. Following the generated design idea: \textit{``it blends global and smooth gradients using the cosine similarity confidence when sign consistency fails, instead of selecting the one with smaller magnitude''.} As a result, this makes the added mode a more aggressive gradient-conditioned Planar-like surface, in contrast to the boundary-filtering refinement found in the replacement case.
	
	%\section{Discussion and Conclusion}
	\section{Conclusion and Discussion}
	In this study, we present preliminary experiments on whether LLMs are capable of designing coding tools. The observation is encouraging: using a generation-and-evaluation loop and straightforward prompt constraints that keep the predictor lightweight and codec-compatible, LLMs can discover meaningful and competitive Planar predictors under nontrivial coding settings. Specifically, for the VVenC \textit{faster} preset, we demonstrate that novel Planar predictors with distinct mechanisms can be obtained by controlling the selection of the top-16 candidate pool with $\lambda$. Notably, despite being evolved on natural-content sequences, the predictors still generalize reasonably well to screen content (Class F), suggesting the benefit from a diverse training set. Meanwhile, even on the highly saturated ECM platform, further bitrate savings remain achievable through both mode replacement and mode insertion under a restricted low-resolution setting.
	
	However, the current workflow still faces a clear obstacle: evaluation is computationally too expensive. Finding an effective predictor depends heavily on exploring a large function space. For example, in our ECM experiment, nearly 2000 new candidates are evaluated before a useful predictor is observed. With parallel evaluation on 256 logic cores (AMD EPYC 9754 processors), one iteration still takes about one hour. More importantly, our results show that a predictor found at low resolution does not necessarily transfer well to higher resolutions. Similarly, a predictor effective under the simpler VVenC \textit{faster} preset may fail under more saturated tool configurations such as \textit{slower}. Therefore, a low-complexity in-domain exploration strategy is critical. For prediction tools, existing expert workflows often use residual signals to pre-select candidates before full encoding, and similar procedures may also benefit the LLM workflow. There are also several directions for improvement. First, the feedback context to the LLM can be richer. The current loop mainly returns the code, design idea, and sequence-level BD-rate of historical candidates, while richer information such as residual maps, content features, mode competition statistics, and failure cases could make the LLM more comprehensively informed. Second, coding tool design inherently couples tool logic, syntax, and RDO algorithms. An effective joint optimization workflow would further extend the scope of LLM-based coding tool design. Looking forward, if such a workflow becomes substantially more efficient and reliable, it may provide a new source of candidate tools for future standardization exploration.
	%Conceivably, if such workflow becomes substantially more efficient and reliable, it is plausible that future standardization could adopted coding tools generated end-to-end by LLMs.

	%This opens a concrete path toward automatic tool design, where future work should focus on cheaper evaluation, richer feedback, and joint optimization of prediction, syntax, and encoder-side decisions.
	%
	%It is conceivable to extend to other types of coding tools.
	%Over the years, the coding tool design has remained one of the most challenging domains in video coding research. If the loop studied here becomes substantially more efficient and reliable, it is plausible that future standardization efforts could witness coding tools generated end-to-end by LLMs.
	
	\clearpage
	\bibliographystyle{IEEEtran}
	\bibliography{refs}
	
\end{document}